\documentclass[11pt]{article}

\def\titel{Quantum Space-Time and Tetrads}
\def\be{\begin{equation}}
\def\ee{\end{equation}}
\def\ba{\begin{array}}
\def\ea{\end{array}}
\def\bea{\begin{eqnarray}}
\def\eea{\end{eqnarray}}

\usepackage{amssymb}

\def\set1{ \rm 1\hspace{-0.24em} l }
\def\setR{ {\mathbb I}\hspace{-.34em}{\rm R} }
\def\setC{ \rm \hspace{0.2em} \rule[0.1ex]{0.1ex}{0.63em}\hspace{-0.27em}{C} }
\def\setS{\mathbb S}

\newtheorem{conclu}{Conclusion}

\begin{document}
\sloppy

\thispagestyle{empty}

\begin{center}

{\Large\bf \titel}\footnote{\em Submitted to
International Journal of Theoretical Physics}

\vspace{5mm}

{\large Holger Lyre}\footnote{\em
Institute of Philosophy,
Ruhr-University Bochum,
D-44780 Bochum,
FRG,\\
email: holger.lyre@rz.ruhr-uni-bochum.de
}

\vspace{5mm}

{\large November 1996}

\vspace{9mm}

\begin{abstract}

The description of space-time in a quantum theoretical
framework must be considered as a fundamental problem in physics.
Most attempts start with an already given classical space-time - then
the quantization is done.
In contrast to this the central assumption in this paper is not to
start with space-time, but to derive it from some more abstract
presuppositions like this is done in {\sc von Weizs\"{a}cker}'s
{\em quantum theory of ur-alternatives}.
Mathematically the transition from a manifold with spin structure
to a manifold with four real space-time coordinates has to be
considered. The suggestion is made that this transition can be well
described by using a tetradial formalism which appears to be the
most natural connection between ur-spinors and real four-vectors.

\end{abstract}

\end{center}

\section{Introduction}

This paper is a speculative one,
it deals with some perspectives in describing space-time from a more
fundamental point of view: from an underlying ''world'' of spinors.
Our starting point is the so called
{\em quantum theory of ur-alternatives}
which is based on the physical and philosophical considerations of
{\sc Carl Friedrich von Weizs\"{a}cker} \cite{cfw85}.
The idea is that the primary substratum in the world is been given by urs:
the simplest objects in quantum theory.
Conceptually they represent nothing further than
{\em 1 bit of potential information} \cite{lyre95}.
Thus, they could be called quantum bits (in modern
quantum information theory sometimes called ''qubits'').
It is mathematically trivial that each physical object
in quantum theory may be embedded in a tensor product of urs.
Let us consider the essential symmetry group of urs which is
\be
U(2) = SU(2) \otimes U(1) \ \sim \ \setS^3 \times \setS^1 .
\ee
In ur-theory the central assumption is that the group-manifold
$\setS^3 \times \setR^+$,
which is associated to the universal symmetry group of urs $U(2)$,
has to be looked upon as a model of our global cosmic space and time
(we take $\setR^+$ instead of $\setS^1$ to describe time because
of our philosophical motivation: we have to presuppose the difference
between past and future which is essential in empirical science).

What is the argument for that astonishing assumption?
As far as any empirical object is build up from urs the
symmetry properties of urs have to be the symmetry properties
of any empirical objects, but the latter ones are essentially the
space-time symmetry properties.
Thus, space and time are not just given as a universal background in
physics but reflect the fundamental structure of ur-spinors.
A familiar assumption on deriving space-time from spinors had been given
by {\sc Roger Penrose} in his twistor theory \cite{penrose+rindler84}.
{\sc David Finkelstein} gave a suggestive name to these kinds of
programms: ''spinorism'' \cite{finkelstein94}.

To summarize our starting point:

\begin{quote}
{\bf Basic Assumption}\hfill\\
{\em
Ur-spinors are to be considered as the fundamental physical entities.
The quantum theory of urs leads to the universal symmetry group
\[ U(2) = SU(2) \otimes U(1) . \]
Since physical objects are build up from urs the symmetry group of
urs (conceived as a homogeneous space of the group itself) gives
a model of global space-time
\[
\setS^3 \times \setS^1 \to \setS^3 \times \setR^+ .
\]
}
\end{quote}

\begin{quote}\begin{conclu}\hfill\\
\label{conclu1}
The three-dimensionality of position-space and the
one-dimensionality of time are derived.
The first approximation of global position-space,
i.e., the cosmic model $\setS^3$,
is characterized as a maximal symmetric space which allows a
{\sc Killing}-group with 6 parameters which is given by $SO(4)$.
The curvature of our cosmos is $k=1$.
In the static case this is an {\sc Einstein}-cosmos.
\end{conclu}\end{quote}

\section{Ur-Theory in the Global Einstein-Cosmos}

\subsection{Ur-Spinors}

According to conclusion \ref{conclu1} urs are to be considered
as non-localized functions in the global {\sc Einstein}-cosmos $\setS^3$.
We choose a special parametrization to represent them.
A general element of $U(2)$ can be written as
\be
\label{su2}
A = U e^{i\varphi}
  = \left( \begin{array}{cc} a \ & \ b \\ -b^* \ & \ a^* \end{array} \right)
    e^{i\varphi} , \qquad a, b \in \setC, \ \varphi \in \setR .
\ee
From the unitarity condition
\be
A^+ A = U^+ U = \set1_{2 \times 2}
\ee
we have
\be
det \ U = a^* a + b^* b = 1 .
\ee
With $a=w+iz$ and $b=y+ix$ this is equivalent to
\be
\label{wxyz}
w^2+x^2+y^2+z^2=1 ,
\ee
i.e., a representation of $\setS^3$. Thus, urs are non-localized
functions on this group manifold. E.g., the two columns of
(\ref{su2}) represent the ur-spinors $u^A$ and $v^A$ with components
\be\label{ur-spinors}
u^1=a, \quad u^2=-b^*, \quad v^1=b, \quad v^2=a^* .
\ee
In spinor space we use the metric
\be
\big( \epsilon_{AB} \big) = \big( \epsilon^{AB} \big)
= \left( \begin{array}{cc} 0 \ & \ 1 \\ -1 \ & \ 0 \end{array} \right)
\ee
which acts like
\be
u_A = \epsilon_{AB} u^B , \qquad
u^A = \epsilon^{BA} u_B =u_B \epsilon^{BA} ,
\ee
and therefore the covariant components of (\ref{ur-spinors}) are
\be
u_1 =   u^2 = -b^*, \quad
u_2 = - u^1 = -a,   \quad
v_1 =   v^2 =  a^*, \quad
v_2   - v^1 = -b .
\ee
Hence the ur-spinors are orthogonal
\be
\label{orthogonal}
u_A u^A = v_A v^A = 0
\ee
and fulfill the conditions
\be
\label{ur_dyad}
v_A u^A = - u_A v^A = 1 .
\ee
Thus, the ur-spinorial system represents a {\em dyad}.

\subsection{Ur-Tetrads}

The main motivation of ur-theory is the foundation of space-time
structure from the symmetry of urs.
Therefore, we have to look for an appropriate mathematical tool to
express this.
Because of the equivalence of a spinorial dyad to a tensorial tetrad,
such a tool is been properly given by the tetradial formalism.
A tetrad (vierbein) must be looked upon as a spatial reference frame
which is represented by a system of four linear independent 4-vectors
$t_{\mu}^{(\alpha)}$ numbered by the index put in brackets.
The metric tensor of space-time is given by
\be
\label{tetrad_metric}
g_{\mu\nu} = g_{(\alpha)(\beta)} \ t_{\mu}^{(\alpha)} t_{\nu}^{(\beta)} ,
\ee
whereas $g_{(\alpha)(\beta)}$ obeys the condition
\be
g_{(\alpha)(\beta)} g^{(\beta)(\gamma)}
= g_{(\alpha)}^{\ (\gamma)} = \delta_{(\alpha)}^{\ (\gamma)}
\ee
and defines the lower tetradial indices
\be
t_{(\alpha) \mu} = g_{(\alpha)(\beta)} \ t_{\mu}^{(\beta)} .
\ee

A special tetrad is given by using null vectors, i.e.,
$t^{(\alpha)}_{\mu} t^{\mu (\alpha)} = 0$.
It turns out that the relations (\ref{orthogonal}) and (\ref{ur_dyad})
are suitable to define four lightlike 4-vectors in the following way
\be\label{lightlike}
\begin{array}{ccc}
l^{\mu}   = \frac{1}{\sqrt{2}} \ \sigma^{\mu}_{\dot A B} v^{\dot A} u^B ,
& \qquad &
m^{\mu}   = \frac{1}{\sqrt{2}} \ \sigma^{\mu}_{\dot A B} v^{\dot A} v^B ,
\\ \\
l^{* \mu} = \frac{1}{\sqrt{2}} \ \sigma^{\mu}_{\dot A B} u^{\dot A} v^B ,
& \qquad &
n^{\mu}   = \frac{1}{\sqrt{2}} \ \sigma^{\mu}_{\dot A B} u^{\dot A} u^B ,
\end{array}
\ee
whereas the dotted indices denote the complex conjugate spinor
components and $\sigma^{\mu}$ are the {\sc Pauli}-matrices.
The vectors (\ref{lightlike}) fullfil the conditions
\be
l_{\mu} l^{* \mu} = 1, \qquad m_{\mu} n^{\mu} = -1, \qquad 0 \ else.
\ee
Thus, from the spinorial dyad (\ref{ur_dyad}) a null tetrad
\be
\label{null_tetrad}
t_\mu^{(\alpha)}=(l_\mu, l^*_\mu, m_\mu, n_\mu)
\ee
can be gained such that the {\sc Minkowski}an metric
$\eta_{\mu\nu} = diag(-1,1,1,1)$ turns out to be
\be
\label{metric}
\eta_{\mu\nu} = l_{\mu} l^*_{\nu} + l^*_{\mu} l_{\nu} - m_{\mu} n_{\nu} - n_{\mu} m_{\nu}
\ee
with
\be
g_{(\alpha)(\beta)} = \left( \begin{array}{cccc}
0 & 1 & & \\ 1 & 0 & & \\ & & 0 & -1 \\ & & -1 & 0 \\
 \end{array} \right) .
\ee
With (\ref{ur-spinors}) we additionally find
\be
\label{m_n}
m_0=n_0=1, \qquad m_k=-n_k .
\ee
The relations (\ref{lightlike}), (\ref{null_tetrad}) and
(\ref{metric}) are based on $SL(2,\setC)$ as invariance group of
ur-spinors, whereas (\ref{m_n}) is a consequence of the
$SU(2)$-representation according to (\ref{su2}), (\ref{ur-spinors}).

\begin{quote}\begin{conclu}\hfill\\
The ur-spinorial dyad (\ref{ur_dyad}) is in a \underline{in a natural way}
associated with a null-tetradial reference frame
(\ref{lightlike}), (\ref{null_tetrad}).
Relation (\ref{metric}) can be understood as the derivation of the
pseudo-{\sc{Euclide}}an structure of space-time from ur-tetrads.
\end{conclu}\end{quote}

{\sc Th. G\"ornitz} pointed out
that the urtheoretical cosmos $\setS^3$ is expanding with
\be
\label{RofT}
R(T) = R(0) + c \cdot T ,
\ee
whereas $R$ is the curvature radius of $\setS^3$, $T$ the cosmic
epoch, and $c$, presumably, the velocity of light \cite{goernitz88b}.
Thus, ur-theory leads to a
{\sc{Friedmann}}-{\sc{Robertson}}-{\sc{Walker}}-cosmos with (\ref{RofT})
in agreement with the {\sc Einstein}-equations.

\begin{quote}\begin{conclu}\hfill\\
From the ur-spinorial invariance group $SL(2,\setC) \sim SO(1,3)$
and from the postulate of a universal limiting velocity $c$
we can derive the full special relativity theory
from the quantum theory of urs.
\end{conclu}\end{quote}

Because the vectors (\ref{lightlike}) are complex,
we choose real linear combinations
(compare {\sc Penrose} and {\sc Rindler}
\cite[Vol. II, p. 120]{penrose+rindler84}).
They are explicitly given by
\bea
\label{tetrad_reell0}
t^{\mu} &=& \frac{1}{\sqrt{2}} \ \left( m^{\mu}+n^{\mu} \right)
= \left( \begin{array}{c} 1 \\ 0 \\ 0 \\ 0 \end{array} \right) , \\
\label{tetrad_reell1}
z^{\mu} &=& \frac{1}{\sqrt{2}} \ \left( m^{\mu}-n^{\mu} \right)
= \left( \begin{array}{c}
0 \\ ab+a^*b^* \\ i(ab-a^*b^*) \\ bb^*-aa^* \end{array} \right)
= \left( \begin{array}{c}
0 \\ 2(wy-xz) \\ -2(wx+yz) \\ x^2+y^2-w^2-z^2 \end{array} \right) , \\
\label{tetrad_reell2}
x^{\mu} &=& \frac{1}{\sqrt{2}} \ \left( l^{\mu}+l^{* \mu} \right)
= \frac{1}{2} \ \left( \begin{array}{c}
0 \\ a^2-b^2+a^{*2}-b^{*2} \\ i(a^2-b^2-a^{*2}+b^{*2}) \\
2(ab^*+a^*b) \end{array} \right)
= \left( \begin{array}{c}
0 \\ x^2-y^2+w^2-z^2 \\ 2(xy-wz) \\ 2(wy+xz) \end{array} \right) , \\
\label{tetrad_reell3}
y^{\mu} &=& \frac{i}{\sqrt{2}} \ \left( l^{\mu}-l^{* \mu} \right)
= \frac{1}{2} \ \left( \begin{array}{c}
0 \\ i(a^2+b^2-a^{*2}-b^{*2}) \\ -a^2-b^2-a^{*2}-b^{*2} \\
2i(ab^*-a^*b) \end{array} \right)
= \left( \begin{array}{c}
0 \\ -2(xy+wz) \\ x^2-y^2-w^2+z^2 \\ 2(wx-yz) \end{array} \right) .
\eea

The system (\ref{tetrad_reell0}) - (\ref{tetrad_reell3}) is
not a null-tetrad. The dreibein-frame $(\vec x, \vec y, \vec z)$
is a tangent-space at the point $(w,x,y,z)=(1,0,0,0)$ of $\setS^3$
according to (\ref{wxyz}) such that (\ref{su2}) becomes
the identity matrix $\set1_{2 \times 2}$.
The dreibein (\ref{tetrad_reell1}) - (\ref{tetrad_reell3})
must be $SO(4)$-rotated in order to get
a tangent-space at each point of $\setS^3$.

\subsection{The Quantized Ur-Tetrad}

Up to this point we have used urs as spinorial wavefunctions, i.e.,
we considered an ur as the first step of quantization of a simple
alternative. The second quantization is done by the replacement
$u_r \to \hat a_r$ and $u_r^* \to \hat a_r^+$
and the {\sc Bose} commutation relations
\be
\label{CR}
\big[ \hat a_r, \hat a_s^+ \big] = \delta_{rs} , \qquad
\big[ \hat a_r, \hat a_s   \big] = \big[ \hat a_r^+, \hat a_s^+ \big] = 0 .
\ee
Thus, we get a quantum field theory of urs, i.e., a many-ur~-~theory
with a variable number of urs. Consequently from (\ref{CR}) the
quantization of the ur-tetrad
(\ref{tetrad_reell0}) - (\ref{tetrad_reell3}) follows.
We use a special choise of the components of the bispinorial ur
${u \choose u^*}$, i.e., $u_r$ with $r$=1...4 ($u^*$ denotes an anti-ur),
which belongs to a representation of $SL(2,\setC) \oplus SL^*(2,\setC)$
\renewcommand{\arraystretch}{1.5}
\be
\ba{cccc}
u_1   = a   \ e^{ i \varphi} , & u_2 = -b^* \ e^{i \varphi} , &
u_3   = b   \ e^{ i \varphi} , & u_4 =  a^* \ e^{i \varphi} , \\
u_1^* = a^* \ e^{-i \varphi} , & u_2^* = -b \ e^{-i \varphi} , &
u_3^* = b^* \ e^{-i \varphi} , & u_4^* =  a \ e^{-i \varphi} .
\ea
\ee
\renewcommand{\arraystretch}{1}
With the abbreviations
\be
\hat\tau_{rs} = \frac{1}{2} \big\{ \hat a_{r}^{+}, \hat a_{s} \big\}, \qquad
\hat n_{r}    = \hat\tau_{rr} , \qquad
\hat n        = \sum_{r} \hat n_{r}
\ee
we get
\be
\label{ur-tetrad}
\ba{lcl}
\hat t^{\mu} = \pmatrix{ \hat n \cr 0 \cr 0 \cr 0 } ,
& \quad &
\hat z^{\mu} = \frac{1}{2} \pmatrix{ 0 \cr
     - \hat\tau_{12} - \hat\tau_{21} + \hat\tau_{34} + \hat\tau_{43}  \cr
 i (   \hat\tau_{12} - \hat\tau_{21} - \hat\tau_{34} + \hat\tau_{43}) \cr
     - \hat n_1 + \hat n_2 + \hat n_3 - \hat n_4                      } ,
\\ \\
\hat x^{\mu} = \frac{1}{2} \pmatrix{ 0 \cr
        \hat\tau_{14} + \hat\tau_{41} + \hat\tau_{23} + \hat\tau_{32}  \cr
  i ( - \hat\tau_{14} + \hat\tau_{41} - \hat\tau_{32} + \hat\tau_{23}) \cr
        \hat\tau_{13} + \hat\tau_{31} - \hat\tau_{24} - \hat\tau_{42}  }  ,
& \quad &
\hat y^{\mu} = \frac{1}{2} \pmatrix{ 0 \cr
  i ( - \hat\tau_{14} + \hat\tau_{41} - \hat\tau_{23} + \hat\tau_{32}) \cr
      - \hat\tau_{14} - \hat\tau_{41} + \hat\tau_{23} + \hat\tau_{32}  \cr
  i ( - \hat\tau_{13} + \hat\tau_{31} + \hat\tau_{24} - \hat\tau_{42}) }  .
\ea
\ee

Of course, this is just a first very simple version of the quantization
of the global space-time model in terms of a tetradial system
of ur-operators. In the language of quantum gravity
$t_{\mu}^{(\alpha)}$ represents four vector bosons,
i.e., massless ''gravitons'' with spin 1.
It seems that the quantization of urs leads consequently
to a quantization of the ur-tetradial reference frame,
i.e., global space-time.

\section{Outlook}

What can we learn from the quantization of the ur-tetrad
about the fundamental question if space has to be treated
as a continuum? We first look at $t^\mu$ in (\ref{ur-tetrad}).
The cosmic time (the epoch) is correlated with the total number of urs,
i.e., the increase of the number of urs has to be understood as an
expression of time. Consequently, at a certain epoch there will be
only a finite number of urs in the world. This number can be estimated
at about $10^{120}$. {\sc Weizs{\"a}cker} calls this
{\em open finitism} \cite[page 471]{cfw85}.
If we keep the curvature radius $R$ of $\setS^3$ according
to (\ref{RofT}) constant, we find that with passing time
there are more and more alternatives available to divide $R$
into smaller and smaller intervalls.
Equivalently, we could say that our unit sticks to
measure spatial lengths decrease.
But from open finitism, it follows that the procedure of
division, i.e., ''counting'' of ur-alternatives, takes time and,
thus, space is ''continuous'' (i.e., infinite)
only in a \underline{potential} sense.

One hope could be that the proposed tetradial formalism is perhaps
a way to deal with general relativity theory in an ur-theoretic manner.
But we meet with a hard problem at this point:
if we take serious the concept of a space-time manifold which
''exists'' only in a potential sense -- what, then, is the suitable
mathematical description for it? Could it be the tetradial formalism?
These questions have further to be studied.

\end{document}